# In-Situ Growth and Ionic Switching Behavior of Single-Crystalline Silver Iodide Nanoflakes

Amir Parsi[1], Abdulsalam Aji Suleiman[1*], Doruk Pehlivanoğlu[1], Hafiz Muhammad Shakir[2], Emine Yeğin[1], T. Serkan Kasırga[1+]

[1]Bilkent University UNAM – Institute of Materials Science and Nanotechnology, Ankara 06800, Türkiye

[2]Department of Physics, Bilkent University, Ankara 06800, Türkiye

[+]Corresponding author email: kasirga@unam.bilkent.edu.tr

[*]Current Address: Department of Engineering Fundamental Sciences, Sivas University of Science and Technology, Sivas 58000, Türkiye

**Abstract**
Silver iodide (AgI) is a prototypical superionic conductor, undergoing a first-order phase transition at 147 °C that enables rapid ionic transport through its lattice, making it attractive for solid-state ionic devices. However, due to the presence of mobile Ag ions, controlled chemical vapor deposition (CVD) synthesis of high-quality AgI single crystals has remained largely unexplored. Here, we present the controllable synthesis of thin, single-crystalline β-AgI nanoflakes using a home-built CVD setup with real-time optical observation capability, offering insights into their nucleation and growth dynamics. We evaluate the material's environmental stability through temperature-dependent photodegradation and Ag nanoparticle formation induced by electron beam irradiation. Electrical measurements on two-terminal devices with silver contacts demonstrate a remarkable six-order-of-magnitude resistance drop in lateral configurations at elevated temperatures, indicative of switchable ionic conductivity. Additionally, vertical device architectures exhibit clear memristive (resistive switching) characteristics, likely due to the formation of conductive filaments. Our work addresses the key synthesis challenges and highlights the unique electrical properties of thin AgI single crystals, suggesting its potential for innovative devices in unconventional computing, data storage, and advanced neuromorphic systems.
**Keywords:** Single-crystalline AgI, Real-time optical monitoring, Chemical vapor deposition, Ionic conductivity, Resistive switching devices

**Main Text**

## 1. Introduction

Superionic conductors (SICs) are solid-state materials known for their high ionic conductivities, akin to those in molten salts, approximately 1 S/cm [1, 2]. SICs often undergo an order–disorder transition to reach the superionic state, categorized into three classes based on the nature of the transition. Type I SICs, such as AgI, have a distinct first-order phase transition, while type II SICs, like $PbF_2$, show a continuous transition across the phases.  and type III SICs, like the β-alumina, lack a well-defined transition [1]. These SICs possess high ionic conductivity, reflected in a diffusion coefficient D of about $10^{-5} cm^2 s^{-1}$, similar to how fast ions move in liquid water at room temperature [1-3].

AgI has been an archetypal material to study type I SICs due to its peculiar properties and prospective applications [2, 4, 5]. It exhibits three polymorphs: body-centered cubic (α-AgI), hexagonal wurtzite (β-AgI), and cubic zinc blende (γ-AgI) [1, 2, 4, 6-10]. Below the critical temperature ($T_c = 147$ °C), AgI exists in the low-conducting β and γ phases. At $T_c$, it undergoes a sharp solid-state phase transition to the superionic α-AgI. During this transition, the iodine sublattice changes from a hexagonal close-packed (hcp) to a body-centered cubic (bcc) structure, while the silver sublattice shifts from an ordered to a disordered configuration. Below $T_c$, the silver ions are relatively immobile. Above this temperature, however, they exhibit a liquid-like behavior, hopping between tetrahedral sites within the iodine sublattice. This transition is considered to underlie the substantial ionic conductivity (~1 S/cm) observed in the α phase and the sharp, four-fold increase in conductivity at the critical temperature [1, 2, 10]. The conductivity remains relatively stable up to the melting point of 552 °C, after which it decreases slightly [1, 2, 4, 9].

In its three-dimensional (3D) form, AgI typically exhibits a direct band gap (~2.8 eV), making it appealing for various optics applications [11]. Lucking et al. theoretically demonstrated that AgI can retain structural stability even at the monolayer level [12]. Recently, the experimental realization of poly-crystalline AgI thin film was achieved through room-temperature adsorption of molecular iodine onto specific Ag substrates [13, 14]. However, ionic conductivity studies are often limited by the use of polycrystalline samples, where randomly oriented grains and grain boundaries introduce scattering sites that obscure intrinsic transport mechanisms. These structural inhomogeneities complicate the interpretation of ionic transport behavior. In contrast, single-crystalline samples provide well-defined transport pathways, enabling more precise investigations into the fundamental mechanisms of ion conduction. Despite their advantages, synthesizing high-quality single crystals remains challenging, leading to a continued reliance on polycrystalline materials in ion transport studies. To the best of our knowledge, there are no reports on the direct synthesis of thin single-crystalline AgI nanostructures.

To address these challenges, we report the controllable vapor-phase growth of thin, single-crystalline β-AgI using a confined-space CVD approach with real-time optical monitoring. This setup allows direct visualization of nucleation and growth dynamics, providing insights into the nanoscale morphological evolution of AgI. The resulting nanoflakes exhibit uniform shape, high crystallinity, and well-defined thickness. We also assess the material's environmental stability by examining its temperature-dependent photodegradation and the formation of Ag nanoparticles induced by electron beam irradiation. Finally, we demonstrate the impact of ionic transport in Ag-contacted two-terminal devices, revealing a six-order magnitude resistance drop in lateral configurations and memristive I-V characteristics in vertical structures.

## 2. Results and discussion

### 2.1. Real-time observation of AgI growth

AgI nanoflakes are grown using a home-built CVD setup that enables real-time optical observation (RTO) and synthesis control, as schematically depicted in **Figure 1a**. Photographs of our setup are given in **Figure S1.** The details of the RTO-CVD method can be found in earlier reports [15-18]. Small AgI granules are evenly distributed on mica loaded onto a substrate heater, after which a freshly cleaved mica, used as a growth substrate, is placed upside down on the AgI powder-loaded substrate to create a confined space configuration. Initially, the chamber is evacuated to 0.3 mbar and then purged with 100 sccm of high-purity Ar for 5 minutes to remove oxygen from the tube. The temperature is then raised to 400 °C at a ramp rate of 15 °C/min and held for 5 minutes to facilitate the successful growth of AgI crystals. Once the crystals of the desired size are observed optically, the substrate heater is turned off to allow natural cooling to room temperature. Throughout the growth process, the carrier gas flow rate (Ar) is maintained at 50 sccm. **SI Video 1** provides a real-time visualization of the crystal growth process.

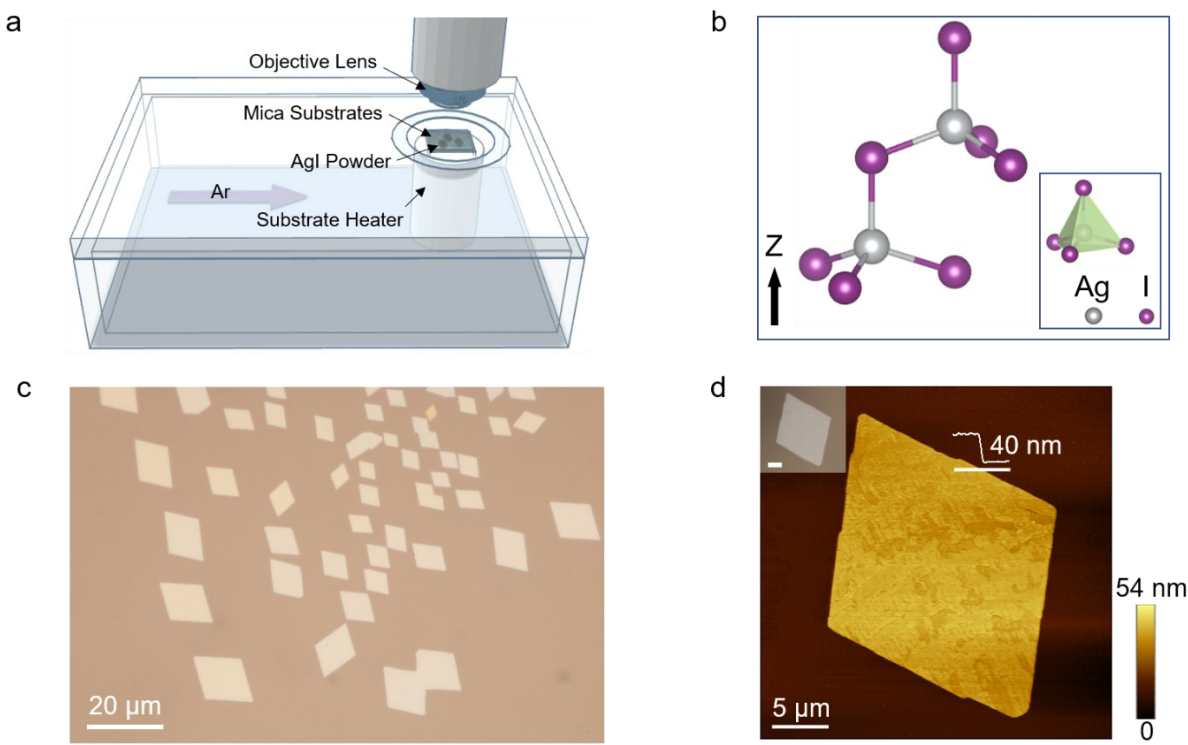

**Figure 1 a.** Schematic of the RTO-CVD chamber used to synthesize AgI crystals. **b.** A schematic representation of the unit cell structure in the hexagonal form of β-AgI. **c.** The optical microscope micrograph of AgI nanoflakes grown on a mica substrate. **d.** AFM height trace map of an as-grown nanoflake with a thickness of 40 nm. The corresponding optical microscope image is shown in the inset. The scale bar is 5 μm.

The growth mechanism of thin, single-crystalline AgI nanoflakes is based on a vapor–solid process, which begins with the sublimation of AgI powder. The vapor-phase species then diffuse toward the upper mica substrate, where they undergo surface adsorption followed by the lateral spreading of adatoms. This sequential process leads to the formation of well-defined AgI nanoflakes on the bottom surface of the top mica substrate.

As demonstrated in **SI Video 1**, gradual color contrast variation, due to optical interference, across different crystals is observed during growth, indicating changes in crystal thickness. The crystals that formed directly on top of the precursor on the growth substrate are larger and thicker, measured by atomic force microscope (AFM) (**Figure S2**). Such a dramatic change in the crystal thickness in the vicinity of the precursor can be attributed to the nonuniform heat distribution around the precursor powder. Regions near the precursor are at higher temperatures as they are in direct contact with the precursor, making them a more favorable site for the adsorption of sublimated precursor molecules. The elevated temperature in this area boosts the adsorption rate of sublimated precursor molecules on the surfaces and edges of the nanoflakes as they grow. A further increase in temperature above 420 °C leads to a decrease in the size and number of crystals.

## 2.2. AgI characterization

We utilized optical, scanning electron, and atomic force microscopy to characterize the synthesized nanoflakes. **Figure 1b** shows the schematic representation of the unit cell for the hexagonal structure of β-AgI, composed of two [AgI$_4$] clusters [11, 13, 19]. **Figure 1c** represents the optical micrograph of typical AgI nanoflakes synthesized on a mica substrate using RTO-CVD. As shown in **Figure 1d**, AFM height trace maps of the AgI crystal reveal a 40 nm thick nanoflake possessing a rhomboidal-shaped morphology with sharp edges. The nanoflakes can grow as large as several tens of micrometers in the lateral direction.

Although the as-synthesized crystals are air-sensitive, they exhibit photo-induced degradation, particularly at elevated temperatures. The photo-induced degradation is not linked to the presence of oxygen, as even in the forming conditions of the RTO-CVD, if the microscope light is kept on during the cool-down period, the illuminated region of the sample would be degraded due to light exposure. **Figure S3** shows post-growth optical micrographs of the areas exposed to light during cool-down. A similar photo-induced degradation is also observed in samples illuminated under an optical microscope heated to 70, 120, 140, and 170 °C (**Figure S4**). The extent of degradation increases significantly above the critical temperature. In contrast, regions not exposed to light show no signs of degradation, even at elevated temperatures. The dots observed on the surface of the crystals are Ag nanoparticles, likely formed due to the photo-induced breaking of bonds between iodine and silver. This bond breaking leads to the removal of the iodine sublattice at high temperatures [19]. These results indicate that AgI crystals should be kept in the dark, especially during high-temperature studies, to prevent Ag island formation.

Additionally, we employed Raman spectroscopy, X-ray diffraction (XRD), X-ray photoelectron spectroscopy (XPS), energy dispersive X-ray spectroscopy (EDX), and high-resolution transmission electron microscopy to characterize the crystals. The XRD analysis (**Figure 2a**) was used to examine the crystallinity and crystal structure of the AgI nanoflakes. The distinct diffraction peaks at 2θ values of 23.8° and 24.6° correspond closely to the characteristic diffraction peaks of the (002) and (101) crystal planes of AgI, confirming its typical hexagonal β-AgI phase (with dimensions of a=4.592 Å, c=7.510 Å, and a space group of $P6_3mc$) indexed to the JCPDS No. 09-0374 [20-22]. The absence of other diffraction peaks may be attributed to the preferential orientation of the crystal surfaces relative to the mica surface. Under ambient conditions, AgI typically exists as a two-phase mixture of the β (hexagonal wurtzite) and γ (cubic zinc blende) phases, depending on the synthesis conditions [4, 10, 23]. As only β-AgI is obtained in its pure form, it could potentially be formed through the compression of tetragonal- or rhombohedral-structured AgI [19]. Two Raman modes were detected at 87 cm$^{-1}$ and 112 cm$^{-1}$ in the AgI samples (**Figure 2b**). These modes correspond to the β-AgI polymorph and are attributed to its E$_2$ and A$_1$ modes, respectively [19, 24].

The elemental compositions of the AgI crystals were analyzed by XPS. The position of elements is calibrated using the 284.8 eV peak of C 1s. The high-resolution XPS scan spectra for Ag 3d and I 3d are presented in **Figure 2c**. The peaks located at 373.8 and 367.8 eV are attributed individually to Ag 3d$^{3/2}$ and Ag 3d$^{5/2}$, indicating the existence of Ag+ in the sample [20, 22].

Furthermore, the binding energies of I 3d$^{5/2}$ and I 3d$^{3/2}$ originating from I- ions within the AgI were observed at 618.8 and 630.3 eV, respectively. EDX analyses are carried out to verify the chemical compositions and purity of the AgI crystals. The EDX spectrum of the AgI crystal is depicted in **Figure 2d**. The EDX spectrum indicates that the AgI nanostructure consists solely of Ag and I, with no impurities present. The distribution of elements was confirmed through EDX elemental mapping. As illustrated in **Figure 2e**, both the Ag and I elements are uniformly dispersed across the nanoflake. The oxygen background is obtained from the mica substrate.

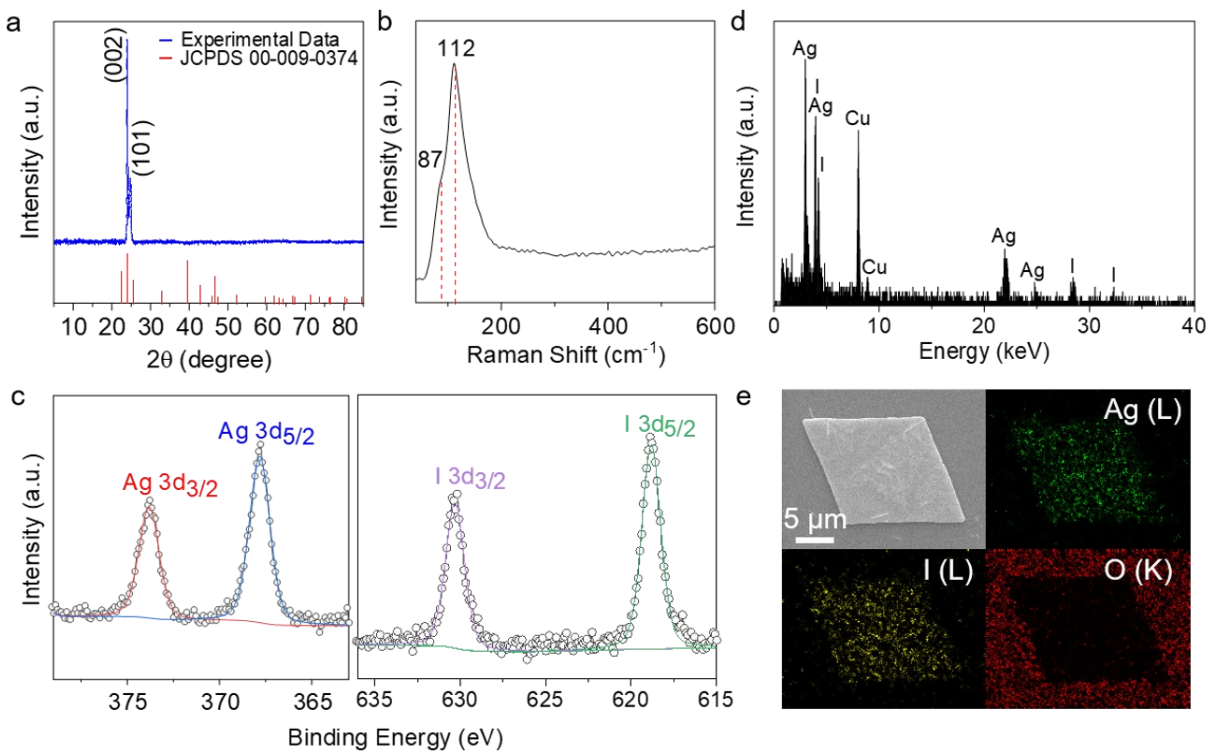

**Figure 2 a.** XRD θ-2θ scan of the AgI crystals at room temperature. **b.** Raman characterization of AgI crystal. **c.** XPS core-level regions of Ag 3d and I 3d, respectively. **d.** EDX spectrum taken in TEM shows that Ag and I are the dominant elements. The traces of Cu are due to the TEM grid. **e.** EDX maps of Ag, I, and O elements.

### 2.3. Formation of Ag nanoparticles upon electron beam irradiation

During SEM imaging and EDX characterization, we observed degradation on the surface of the AgI crystals, accompanied by the formation of new features, similar to those seen in photo-induced degradation of nanoflakes at elevated temperatures. **Figure S5** illustrates the changes caused by exposure to electron beam irradiation. Specifically, **Figure S5b**, **d**, and **f** depict the surface after 1 minute of electron beam exposure. Based on previous studies and our post-e-beam exposure analysis, we concluded that Ag nanoparticles formed on the surface of the AgI crystals [19]. Furthermore, we observed that the number of Ag nanoparticles increased over time or with higher electron beam energy. This phenomenon is attributed to the progressive reduction of Ag$^+$ cations from the bulk to the surface, leading to the degradation of the material and the formation of a composite structure (nAg/Ag$_{1-n}$I).

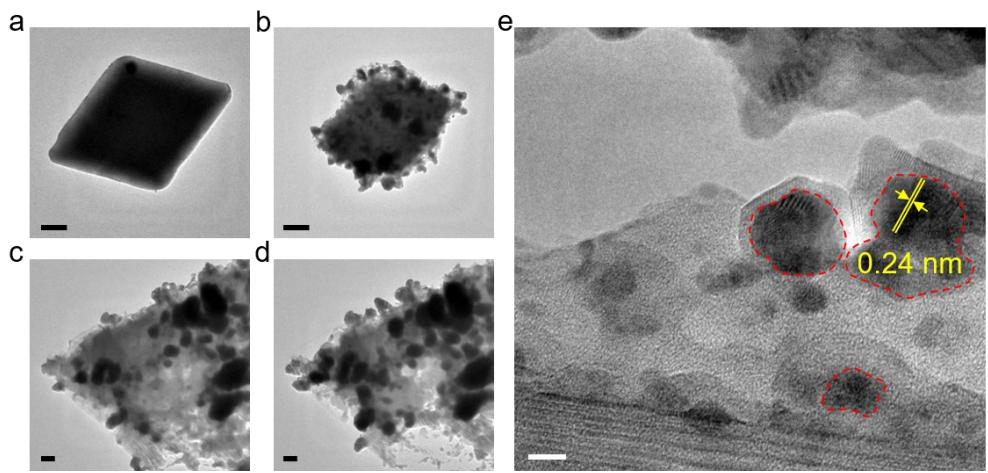

**Figure 3.** TEM analysis of AgI crystals under electron beam irradiation at 200 kV. **(a)** Low-magnification image of an AgI crystal before degradation. **(b)** The same crystal after 20 seconds of irradiation, showing initial formation of Ag nanoparticles. The scale bar in a and b is 0.5 µm. **(c)** A second crystal after 1 minute of irradiation, exhibiting significant degradation and nanoparticle formation (black regions). **(d)** The same crystal from (c) after 5 seconds, highlighting the progression of degradation and expansion of Ag nanoparticles in certain regions. The scale bar in c and d is 0.1 µm. **(e)** High-resolution TEM image of degraded regions, showing dark areas with distinct d-spacing (0.24 nm) corresponding to the (111) planes of the Ag phase (JCPDS No. 04-0783). The red dashed areas show Ag nanoparticles. The scale bar is 5 nm.

**Figure 3** presents the results for transmission electron microscopy (TEM) measurements on AgI crystals. Low-magnification TEM images reveal the formation of Ag nanoparticles on the crystal surface following exposure to electron beam irradiation at 200 kV. The time required to generate these nanoparticles during TEM analysis was much shorter than that in SEM analysis, owing to the significantly higher acceleration voltages used for TEM imaging [25]. A high-resolution TEM image, depicted in **Figure 3e**, highlights darker regions with distinct d-spacing values. Upon measuring these d-spacings, they were identified as corresponding to the (111) crystal planes of the FCC crystal structure of Ag, aligned with JCPDS No. 04-0783 [26]. This observation aligns with findings reported by Longo et al.[25], confirming that electron beam irradiation induces the reduction of Ag within the AgI matrix, resulting in the formation of Ag nanoparticles.

AgI decorated with Ag nanoparticles has shown superior photocatalytic performance [19, 21, 27]. Yuan et al. reported enhanced photocatalytic activity of AgI under a broad solar spectrum, attributed to the presence of coexisting Ag nanoparticles [27]. Similarly, Wang et al. observed the partial formation of Ag nanoparticles alongside AgI, which induced characteristic surface plasmon resonance (SPR) effects and significantly improved the photocatalytic efficiency of Ag/AgI composites [21]. Developing one-pot methods to synthesize Ag/AgI plasmonic photocatalysts remains challenging, highlighting the need for new strategies to grow metallic Ag on AgI for advanced photocatalytic applications.

### 2.4. Device fabrication and electrical measurements

In earlier studies, AgI was employed as the switching medium to potentially construct electrochemical metallization (ECM) memristors, which rely on the formation and dissolution of metallic conductive filaments (CFs) under an electric field. ECM memristors show promise for memory and neuromorphic computing [28, 29]. These devices typically use metal cations from an active electrode, like Ag, to create CFs, switching between low and high resistance states abruptly [26]. While traditional oxide-based ECM memristors face limitations in density and power efficiency, AgI might offer a promising alternative for high-performance, low-power memristors with fast switching and high integration potential. Here, we tested our single-crystalline samples for potential switching applications.

Top-silver-contacted device fabrication started with the transferring of as-grown AgI nanoflakes on an $HfO_2$-coated $SiO_2$/Si substrate using the PMMA-assisted transferring method. In this method, the growth substrate was coated with a PMMA solution and baked at 170 °C for 5 minutes in the dark to prevent crystal degradation. The PMMA-coated sample was then placed in boiling water for 15 minutes and left to cool for 45 minutes. The film was removed from the growth substrate by wedging into water and transferred onto the $HfO_2$-coated $SiO_2$/Si substrate. Finally, the film was dissolved in acetone for 5 minutes, leaving the nanoflakes on the substrate. After dissolving the PMMA film, the nanoflakes are ready to be patterned using the Mask Aligner, followed by thermal evaporation of contacts, Cr (5nm)/Ag (100nm). The inset of **Figure 4a** shows the top-silver-contacted device.

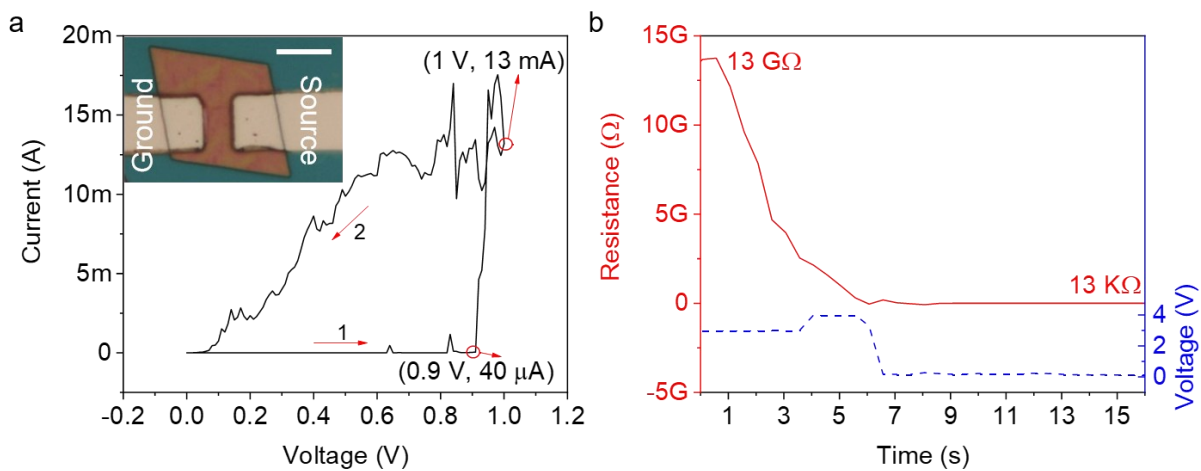

**Figure 4 a.** Output characteristics of a lateral Ag/AgI/Ag device measured at 170 °C, with an optical microscope image of the device shown in the inset. The scale bar is 10 μm. **b.** Resistance as a function of time under varying applied voltages.

We examine how AgI conducts electricity at elevated temperatures (170 °C). The I-V curve reveals a transition from a high resistance state (HRS) to a low resistance state (LRS) at voltages exceeding ~0.9 V (**Figure 4a**). Notably, applying a constant voltage of ~4 V results in a dramatic resistance drop of approximately six orders of magnitude (**Figure 4b**). This behavior is likely due to the formation of conductive Ag nanowires bridging the electrodes, as observed in **Figure S6**. EDX analysis was performed on the device, which was biased positively relative to the ground. The EDX maps reveal a depletion of silver in the source electrode (**Figure S6c**), consistent with $Ag^+$ ions migrating from the biased electrode toward the grounded terminal under the electric field.

When a voltage exceeding a certain threshold is applied, it can induce a liquid-like motion of $Ag^+$ ions within the nanoflake, with the electrodes acting as ion sources or sinks depending on the polarity. The device studied here showed signs of damage and did not exhibit reversible switching behavior. Further investigation on multiple devices is required to evaluate the reproducibility and reversibility of this behavior.

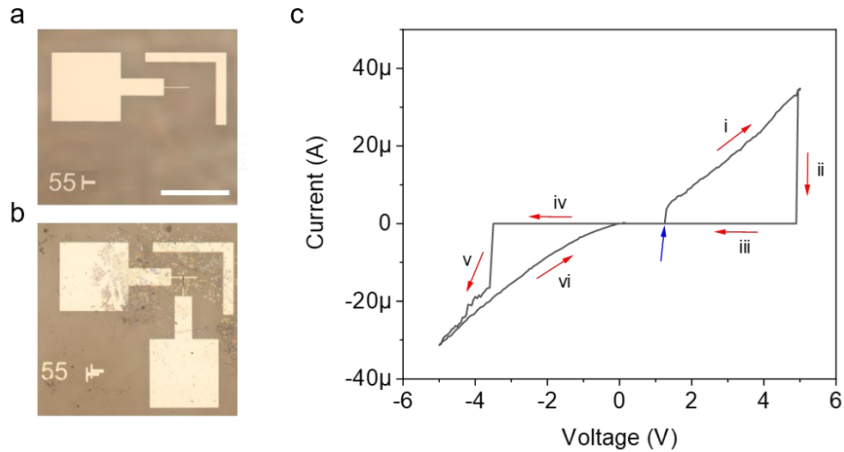

**Figure 5 a.** A 10 nm Cr/50 nm Ag deposited contact on an empty sapphire substrate. **b.** Optical image of the AgI device in a vertical configuration after transferring AgI crystals onto the pre-patterned Ag bottom electrode, followed by patterning and depositing 10 nm Cr/100 nm Ag as the top electrode. The scale bar is 300 µm. **c.** I-V cycles of symmetric AgI devices with Ag contacts in a vertical configuration. The I-V cycle shows large hysteresis windows, exhibiting memristive behavior. The blue arrow points to the forming voltage at 1.25 V. The red arrows indicate the sweep direction.

In addition to lateral configuration devices, AgI devices with a vertical configuration were also fabricated and measured. The process began with patterning bottom electrodes on the substrate via photolithography, followed by thermal evaporation of 10 nm Cr/50 nm Ag and a lift-off process (**Figure 5a**). AgI nanosheets grown on a mica substrate were then transferred using a wet transfer method, similar to the PMMA-assisted technique described earlier, but with an additional step. Instead of being placed directly onto the target substrate, the detached film was first transferred onto PDMS. This intermediate step allowed precise alignment of the nanosheet with the pre-patterned electrodes, ensuring proper placement over the electrode channel. Before bringing the PDMS-supported film into contact with the substrate, a droplet of acetone was applied to the substrate, followed by heating to 50 °C. This facilitated the release of the film from the PDMS onto the substrate. The subsequent steps followed the same procedure as the PMMA-assisted transfer method described earlier. Finally, the top electrode was patterned using a mask aligner after aligning the crystal with the mask pattern, followed by 10 nm Cr/100 nm Ag deposition via thermal evaporation and lift-off. **Figure 5b** illustrates the vertical-configuration device using silver contacts.

**Figure 5c** shows the I-V cycle, where the voltage increases from 0 to 5 V, returns to 0, decreases to -5 V, and then returns to 0 V. Large hysteresis windows are evident in both the positive and negative voltage ranges. Specifically, the device exhibits memristive behavior, with a forming voltage close to 1.25 V. A transition from LRS to HRS occurs at a reset voltage of around 5 V. In the negative range, a transition from HRS to LRS is observed at the set voltage of approximately -3.5 V. This behavior might be attributed to the formation of conductive filaments between the two active silver contacts [26, 30]. Further analysis is required to study the evolution dynamics of conductive filaments.

## 3. Conclusion

In conclusion, we have demonstrated the synthesis of single-crystalline AgI flakes via confined-space CVD under real-time optical observation. We have shown that AgI crystals are highly oriented along the crystal's symmetry axis and are in the β-phase at room temperature. The marked sensitivity to light and electron beam exposure, especially above the phase transition temperature, is studied in depth. Our results show Ag nano-islands forming on the surface and within the crystals upon irradiation. Two-terminal devices of lateral and vertical geometries exhibit a large, ionic-based switching upon application of a voltage bias. The unique combination of facile synthesis, sharp conductivity transitions, and intrinsic memristive capabilities positions thin single-crystalline AgI as a promising candidate for next-generation electronics.

## 4. Acknowledgements

TSK acknowledges funding from TUBITAK under grant #121F366.

## 5. Author Contributions

AP performed the crystal synthesis, characterization, and electrical measurements, with the assistance of AAS, DP, HMS, and EY. TSK conceded the project and guided the experiments. AP wrote the draft, and all authors contributed to the writing of the manuscript.

# In-Situ Growth and Ionic Switching Behavior of Single-Crystalline Silver Iodide Nanoflakes

Amir Parsi[1], Abdulsalam Aji Suleiman[1*], Doruk Pehlivanoğlu[1], Hafiz Muhammad Shakir[2], Emine Yegin[1], T. Serkan Kasırga[1+]

[1]Bilkent University UNAM – Institute of Materials Science and Nanotechnology, Ankara 06800, Türkiye

[2]Department of Physics, Bilkent University, Ankara 06800, Türkiye

[+]Corresponding author email: kasirga@unam.bilkent.edu.tr

*Current Address: Department of Engineering Fundamental Sciences, Sivas University of Science and Technology, Sivas 58000, Türkiye

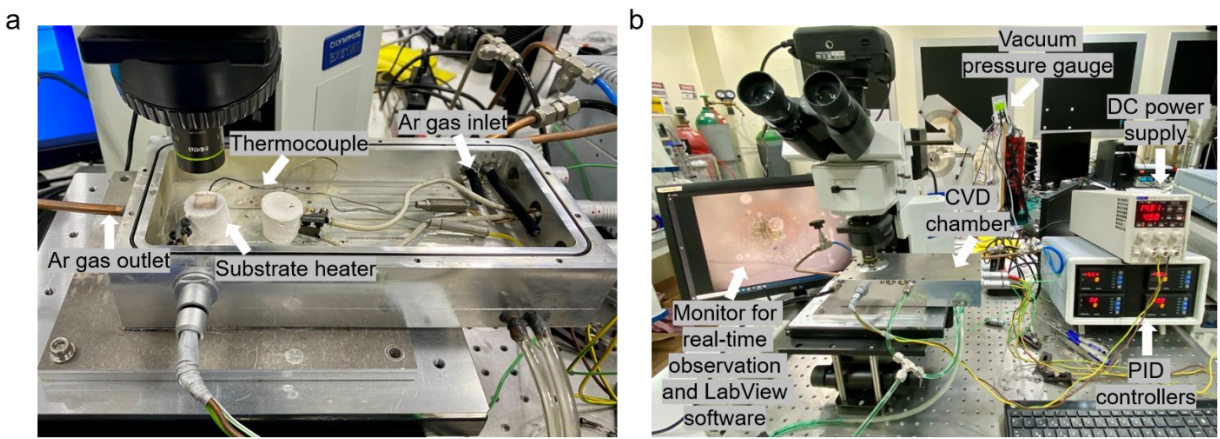

**Figure S1 a.** Close-up view of the custom-made RTO-CVD setup. **b.** The layout of the overall setup. The picture is annotated with key components.

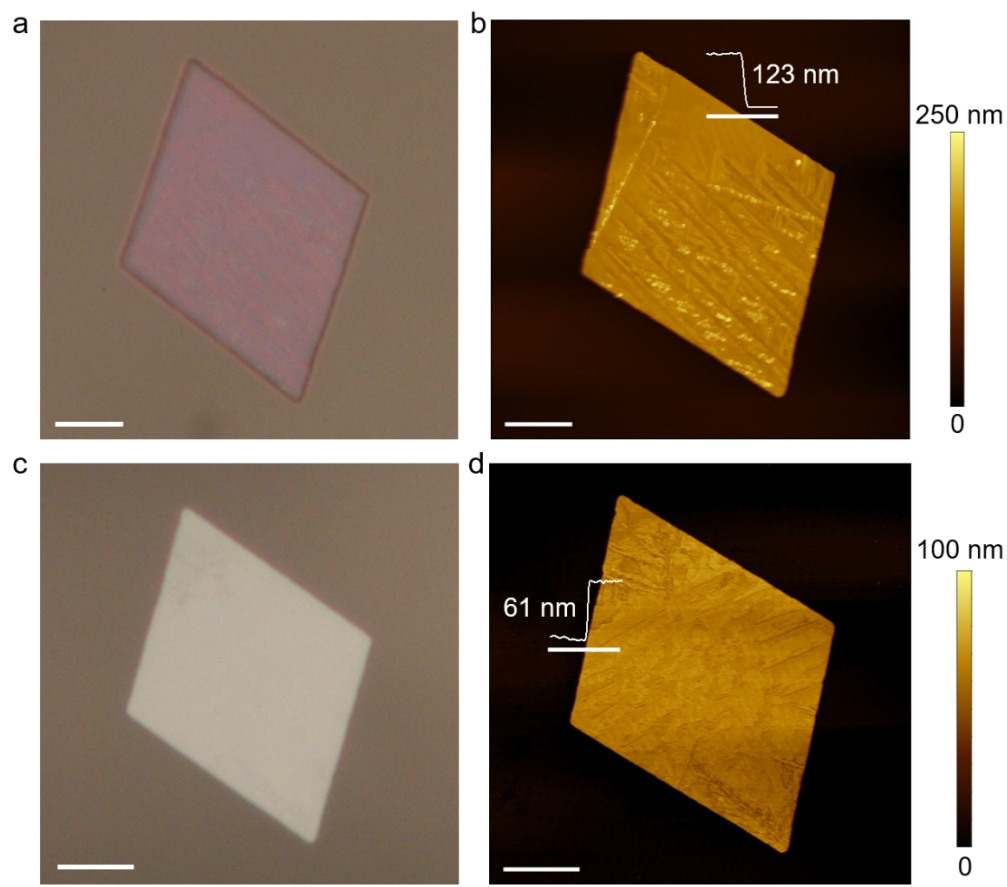

**Figure S2.** Optical micrographs and corresponding AFM height trace maps of AgI nanoflakes illustrating the thickness variation and optical color contrast. The scale bar is 5 μm. **(a)** Crystal formed closer to the precursor, showing a darker optical contrast, and **(b)** its AFM height trace map indicating a thickness of 123 nm. **(c)** Crystal formed farther from the precursor, exhibiting a lighter optical contrast, and **(d)** its AFM height trace map showing a thickness of 61 nm.

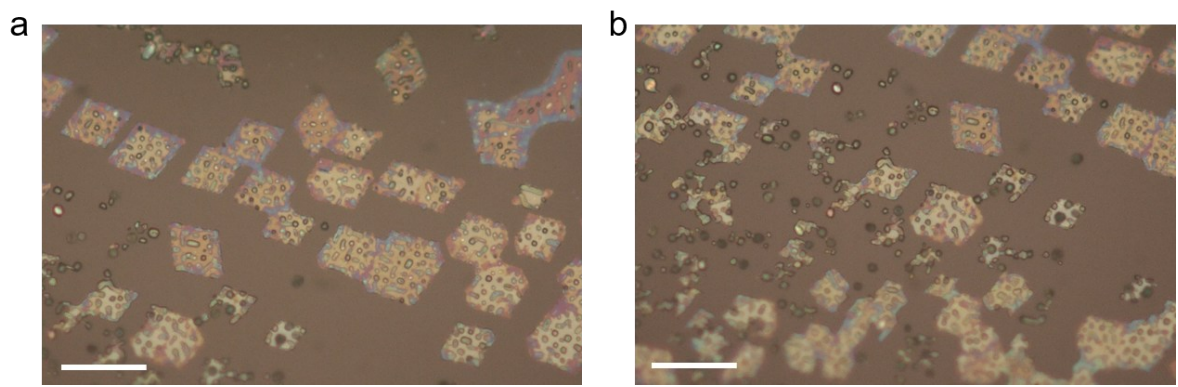

**Figure S3 a** and **b.** Post-growth optical micrographs of the light-exposed area on the substrate. Ag nanoparticles were formed from the irradiation of AgI nanoflakes at high temperatures. The scale bar is 20 μm.

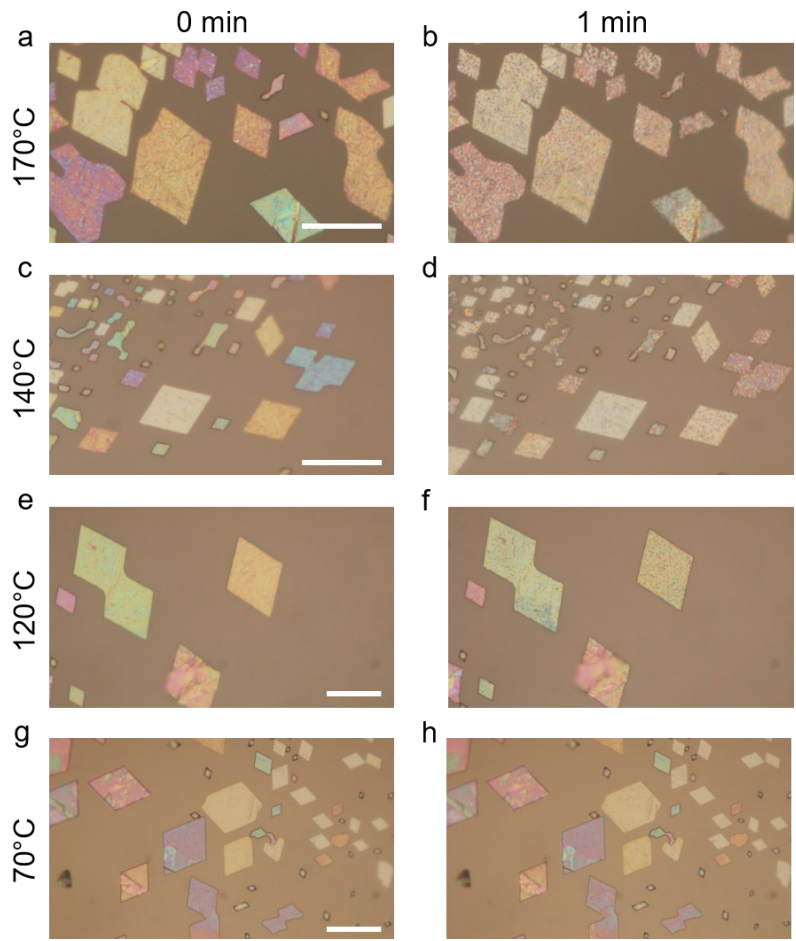

**Figure S4.** Temperature-dependent degradation of AgI crystals under microscope illumination. **(a, b)** Optical micrographs of crystals at 170°C before (a) and after 1 minute of illumination (b). **(c, d)** Crystals at 140°C before (c) and after 1 minute of illumination (d). **(e, f)** Crystals at 120°C before (e) and after 1 minute of illumination (f). **(g, h)** Crystals at 70°C before (g) and after 1 minute of illumination (h), with no significant change observed. The scale bar is 20 μm.

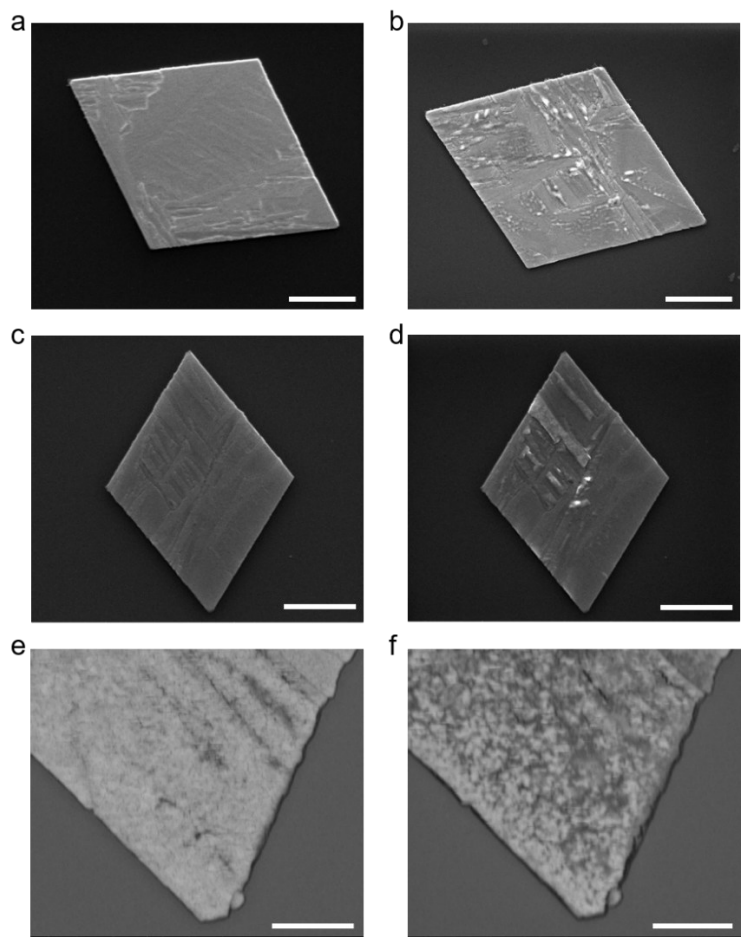

**Figure S5.** Growth of Ag nanoparticles under electron beam irradiation observed during SEM at **(a-d)** 10 kV and **(e, f)** 15 kV. Panels **(b)**, **(d)**, and **(e)** show the surface after 1 minute of electron beam exposure. The scale bar in (a-d) is 5 µm, and in (e) and (f) is 1 µm.

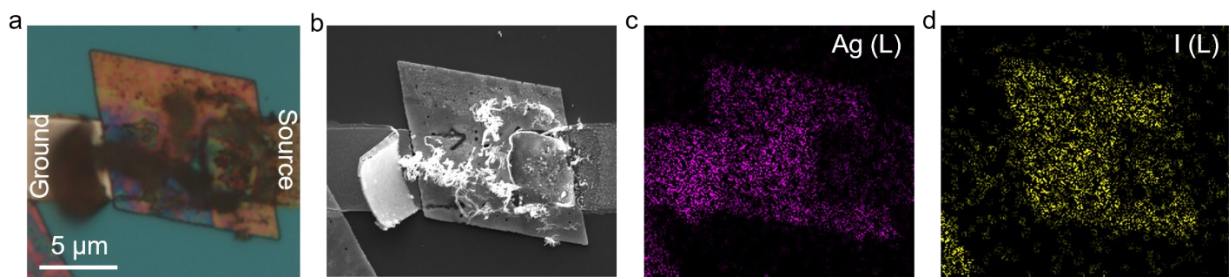

**Figure S6 a.** Optical microscope image of the device following the application of an electric field. **b.** Corresponding SEM image revealing the formation of conductive Ag nanowires between the electrodes. **c-d.** EDX maps displaying the distribution of Ag and I elements after the electric field application, with the source contact showing a depletion of Ag.